# Giant oscillations in a triangular network of one-dimensional states in marginally twisted graphene

S. G. Xu[1,2,*], A. I. Berdyugin[1,*], P. Kumaravadivel[1,2], F. Guinea[1], R. Krishna Kumar[1,2], D. A. Bandurin[1], S. V. Morozov[3], W. Kuang[1], B. Tsim[1,2], S. Liu[4], J. H. Edgar[4], I. V. Grigorieva[1], V. I. Fal'ko[1,2], M. Kim[1,+], A. K. Geim[1,2,+]

[1]School of Physics and Astronomy, University of Manchester, Manchester M13 9PL, UK.
[2]National Graphene Institute, University of Manchester, Manchester M13 9PL, UK.
[3]Institute of Microelectronics Technology and High Purity Materials, Russian Academy of Sciences, Chernogolovka 142432, Russia.
[4]The Tim Taylor Department of Chemical Engineering, Kansas State University, Manhattan, Kansas 66506, USA.

[*]These authors contributed equally to this work.
[+]Correspondence and requests for materials should be addressed to M. K. (minsoo.kim@manchester.ac.uk) or to A. K. G. (geim@manchester.ac.uk).

**At very small twist angles of ~0.1°, bilayer graphene exhibits a strain-accompanied lattice reconstruction that results in submicron-size triangular domains with the standard, Bernal stacking. If the interlayer bias is applied to open an energy gap inside the domain regions making them insulating, such marginally twisted bilayer graphene is expected to remain conductive due to a triangular network of chiral one-dimensional states hosted by domain boundaries. Here we study electron transport through this helical network and report giant Aharonov-Bohm oscillations that reach in amplitude up to 50% of resistivity and persist to temperatures above 100 K. At liquid helium temperatures, the network exhibits another kind of oscillations that appear as a function of carrier density and are accompanied by a sign-changing Hall effect. The latter are attributed to consecutive population of the narrow minibands formed by the network of one-dimensional states inside the gap.**

### Introduction

The electronic properties of graphene superlattices have attracted intense interest[1–4] that was further stimulated by the recent observation of novel many-body states in twisted bilayer graphene (BLG)[2,5–18]. The latter system exhibits qualitative changes with varying the twist angle $\theta$ between the two graphene layers, which is caused by subtle interplay between the interlayer electron hybridization and the periodic crystallographic pattern known as a moiré superlattice[2,5–18]. For small $\theta$, the superlattice period is given by $\lambda = a/[2\sin(\theta/2)] \approx a/\theta$ and is much longer than graphene's lattice constant $a$. The recent interest in twisted BLG has been focused on so-called magic angles (typically, close to $\theta \approx 1.1°$) at which the low-energy superlattice minibands become almost flat[5,6] promoting electron-electron correlation effects and leading to unconventional insulating and superconducting states[8–10.]. At the marginal twist angles, $\theta \ll 1°$, the electronic structure is expected to become qualitatively different from that formed at magic or larger $\theta$ because the BLG superlattice undergoes a strain-accompanied lattice reconstruction such that there appear large (submicron) triangular domains with alternating Bernal (AB and BA) stacking order[11–13]. The domain regions are rather similar to the conventional BLG and, if the displacement field $D$ is applied between the layers, a sizeable energy gap $\delta$ opens in the spectrum[11,19,20], making the domains insulating[19]. Under these conditions, marginally twisted graphene (MTG) bilayers may still remain electrically conductive because walls between AB and BA domains allow one-dimensional (1D) chiral states[11–18,21–24] (Fig. 1a). For an AB/BA domain wall, there are four (2 spins and 2 valleys) gapless 1D states on each side. They propagate in opposite directions for different valleys and split apart at the superlattice's vertices where the stacking changes into AA (Fig. 1a). The unit block for this 2D network is an equilateral triangle with the area $A = \frac{\sqrt{3}}{4}\lambda^2$, half the size of the superlattice unit cell that includes both AB and BA domains.

In this Communication, we study the electronic properties of MTG and report exceptionally strong Aharonov-Bohm oscillations[25] arising from electron interference along the triangular loops forming the chiral network. Yet another type of oscillations is observed in MTG's resistivity as a function of carrier density, which indicates the presence of multiple electronic minibands inside the gap. Our work shows that marginally twisted BLG is markedly distinct from other 2D electronic systems, including BLG at larger twist angles, and offers a fascinating venue for further research.

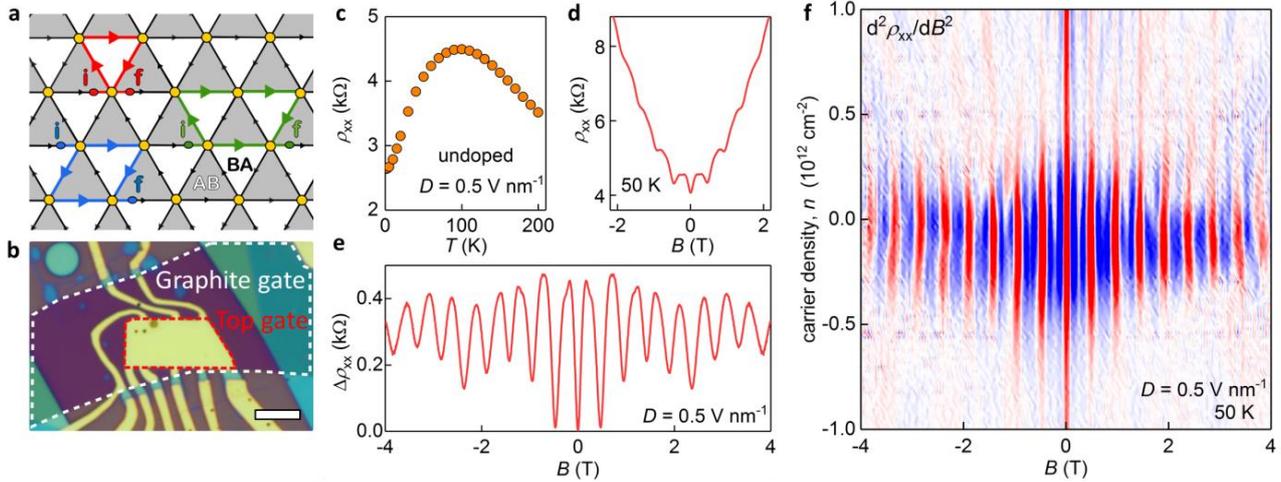

**Figure 1| Aharonov-Bohm oscillations in marginally twisted bilayer graphene. a**, *Schematic of MTG: Its superlattice forms a triangular network of AB/BA domain walls[11–13]. The white and grey areas denote domains with the Bernal stacking; yellow circles – regions with AA stacking. Black arrows show the propagation direction for electrons in one of the valleys; colored arrows are examples of trajectories encircling one (red), two (blue) and three (green) triangular domains, which are responsible for different frequencies in Aharonov-Bohm oscillations. For example, the 2$^{nd}$ harmonic arises from interference between electrons propagating along the trajectories indicated by blue arrows (starting points are marked by 'i'; finishing by 'f').* **b**, *Optical micrograph of one of the studied devices. The graphite and top gates are indicated by color-coded dashed curves. The bright yellow regions are Cr/Au contacts to graphene and the top gate. Scale bar, 5 μm. The device exhibited carrier mobility of ~$10^4$ cm$^2$ V$^{-1}$ s$^{-1}$ at n ≈ $10^{12}$ cm$^{-2}$, which is typical for small twist angles[12,16] where irregularity in positions of domain walls probably causes additional scattering.* **c,** *Temperature dependence of undoped MTG with the energy gap of ~50 meV induced by interlayer bias.* **d,** *Magnetoresistance at the NP for the same device at 50 K.* **e,** *Same as (d) but a monotonic background is subtracted for clarity.* **f,** *Same Aharonov-Bohm oscillations for different doping. Instead of subtracting the magnetoresistance for each n as in (e), we plot $d^2\rho_{xx}(B)/dB^2$ which removes the smooth background without shifting positions of oscillations' extrema. Blue-to-red scale, ±4 kOhm T$^{-2}$.*

## Results

**Experimental devices.** Our devices were made from MTG that was prepared following the procedures developed in ref.[26]. In short, a monolayer graphene crystal was teared into two parts that were placed on top of each other by parallel transfer accompanied by small rotation. In our case the rotation angle $\theta$ was set close to zero (nominally, 0 to 0.1°). Subsequent transport measurements (see below) showed that the resulting bilayers exhibited twist angles of ≤ 0.25°. The MTG crystals were encapsulated in hexagonal boron nitride to improve their electronic quality and, using lithography techniques, shaped into dual-gated Hall bar devices such as shown Fig. 1b (Supplementary Note 1). The top gate was the standard metal-film electrode whereas the bottom gate was thin graphite, which further improved devices' electronic quality and reduced charge inhomogeneity. Four MTG devices were studied in detail, all exhibiting similar behavior. Below we focus on the results obtained for a device with $\theta \approx 0.10°$ (Supplementary Note 2) which had the highest uniformity, as witnessed from practically the same magnetotransport characteristics observed using different contact configurations. For completeness, examples of the behavior observed for other MTG devices are provided in Supplementary Note 3.

**1D conductive network.** To study electron transport through the expected network created by AB/BA

domain walls, we applied the displacement field $D$ using the top and bottom gates (Supplementary Note 1), which opened an energy gap in the Bernal-stacked regions[11,19,20]. For $D = 0.5$ V nm$^{-1}$ (achievable without a risk of damaging the devices), the gap $\delta$ inside AB and BA regions should be[20] ~ 50 meV so that they become highly insulating at temperatures $T \leq 50$ K as known from the experiments on standard BLG[19]. In contrast, our MTG devices exhibited a distinctly metallic behavior such that longitudinal resistivity $\rho_{xx}$ decreased with decreasing $T$, reaching a few kOhms at liquid-helium $T$ (Fig. 1c). This shows that, at the charge neutrality point (NP), MTG hosts a metallic system (Supplementary Note 5), in contrast to Bernal-stacked BLG and in agreement with the expected 1D transport along AB/BA walls. At $T > 100$ K, the temperature behavior changed so that $\rho_{xx}$ decreased with $T$ (Fig. 1c). The latter observation is attributed to thermally activated carriers in the gapped AB and BA regions, similar to the case of standard BLG[19]. The insulating behavior could also be suppressed by field-effect doping. For $D = 0.5$ V nm$^{-1}$, it typically required carrier densities $n$ above $\pm 3$–$5 \times 10^{11}$ cm$^{-2}$ to start populating the conduction and valence bands, as seen from a drop in $\rho_{xx}$ (Supplementary Note 4).

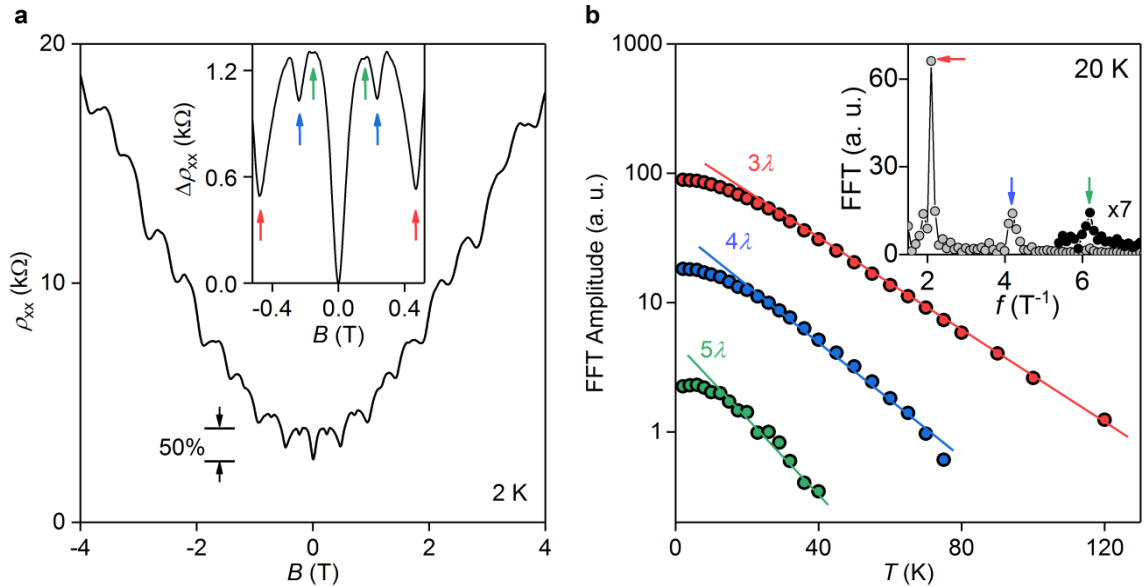

**Figure 2| High-order harmonics in the Aharonov-Bohm oscillations. a**, *Typical magnetoresistance curves near the main NP (T = 2 K, D = 0.5 V nm$^{-1}$). Insert: Zoom-in close to zero B. The arrows mark the minima that correspond to $\phi_0$ piercing one (red), two (blue) and three (green) unit cells shown in Fig. 1a.* **b**, *Temperature dependence for the three oscillation frequencies (symbols). Solid lines: Guides to the eye indicate that the slopes for the 2$^{nd}$ and 3$^{rd}$ harmonics are, respectively, ~ 4/3 and 5/3 times steeper than for the main frequency. Insert: Example of our fast-Fourier-transform (FFT) analysis (grey circles). The peaks are marked using the same color coding as in (a). The third-harmonic peak is magnified for clarity (black symbols).*

**Aharonov-Bohm oscillations.** The network of conducting AB/BA walls revealed itself most clearly in strong magneto-oscillations that were periodic in magnetic field $B$ and observed for all our MTG devices. Examples are shown in Figs. 1d-f and Supplementary Figure 2c. The oscillations developed with increasing $D$ and persisted until AB and BA domains became conductive because of either doping or temperature. For example, at $D = 0.5$ V nm$^{-1}$, this meant $|n|$ up to $5 \times 10^{11}$ cm$^{-2}$ and $T$ up to 120 K as seen in Fig. 1f and Fig. 2, respectively. The periodic-in-$B$ oscillations persisted up to several Tesla where they became overwhelmed by Shubnikov-de Haas oscillations (Supplementary Note 3). We attribute the low-$B$ oscillations to the Aharonov-Bohm effect[25] for electrons propagating along the triangular network of 1D channels hosted by AB/BA walls. Indeed, interference between electronic states propagating along, e.g., the red loop in Fig. 1a is expected to vary periodically with the magnetic flux piercing the domain area $A$. For the particular device in Fig. 1, we found the oscillation period $\Delta B \approx 0.48 \pm 0.02$ T which translates into one flux quantum $\phi_0 = h/e$ per $A \approx$ 0.86±0.04 ×10$^{-10}$ cm$^2$ and yields $\lambda \approx 140$ nm or $\theta \approx 0.10° \pm 3\%$, in agreement with the area found from the position of superlattice NPs (Supplementary Note 2).

As $T$ decreased below 50 K, the Aharonov-Bohm oscillations grew exponentially, reaching more than 1 kOhm in amplitude at liquid-helium $T$, that is, nearly ~50% of zero-$B$ resistivity (Fig. 2a). Furthermore, higher frequency harmonics became visible at low $T$ (inset of Fig. 2a). For quantitative analysis, $\rho_{xx}(B)$ curves for a given $T$ were Fourier-transformed (see the inset of Fig. 2b). The peaks in the Fourier plot reveal the main periodicity $\Delta B \approx 0.48$ T (same as in Fig. 1) plus two fractional periods, $\Delta B/2$ and $\Delta B/3$. The latter correspond to twice and thrice larger areas involved in the interference pattern and can be attributed to the loops such as those indicated by the blue and green arrows in Fig. 1a. This assignment agrees well with the fact that the higher harmonics decayed notably faster with increasing $T$ than the main-frequency oscillations (Fig. 2b), as expected because of the larger circumferences of the blue and green loops. Moreover, the suppression of Aharonov-Bohm oscillations is usually described by the dependence[27] exp(-$L/L_\phi$) where $L$ is the length of the interference loops, and $L_\phi(T)$ is the decoherence length. The color-coded lines in Fig. 2b show that the decay rates for the 1st, 2nd and 3rd frequencies followed the ratios $L/L_\phi$ expected from the circumferences of the three involved loops ($L = 3\lambda$, $4\lambda$ and $5\lambda$, respectively) for a given $L_\phi(T)$.

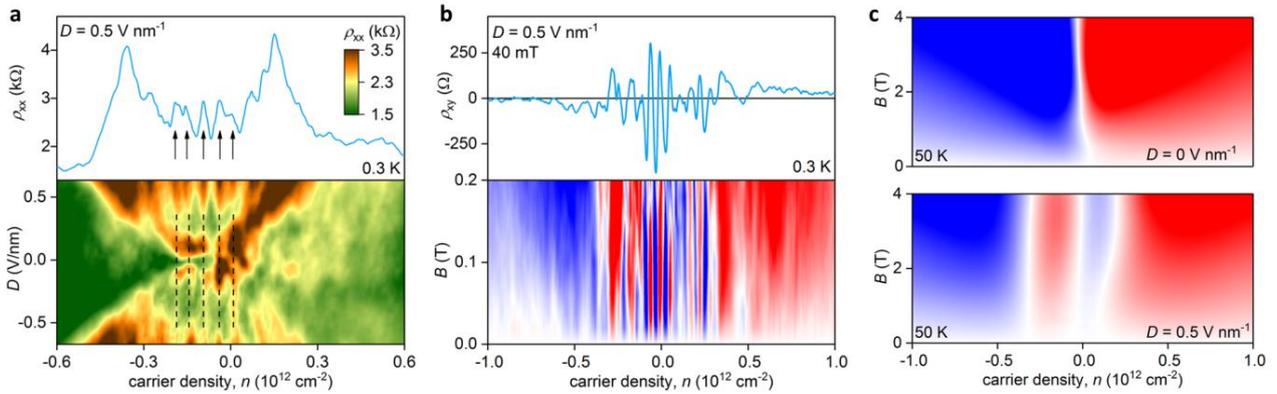

**Figure 3| Minibands in the gapped state of marginally twisted BLG. a,** *Longitudinal resistivity as a function of n and D at 0.3 K. Upper panel: Crosscut of the map at D = +0.5 V nm$^{-1}$.* **b,** *Same oscillatory behavior in Hall resistivity as a function of n and B for the given D. Color scale: ±250 Ohm. Upper panel: Crosscut at B = 40 mT. The arrows in (a) mark some of the maxima in $\rho_{xx}$ which correspond to NPs (zero $\rho_{xy}$) in (b).* **c,** *Same as (b) but measured at 50 K. Upper and lower panels are for D = 0 and 0.5 V nm$^{-1}$, respectively. Color scale: ±3 kOhm.*

**Minibands inside the gap.** We also studied the network's transport properties as a function of doping, keeping $n$ sufficiently low to remain in the insulating state for the AB and BA domains. Figure 3 shows that our MTG devices exhibited strong oscillations in their zero-$B$ resistivity as a function of $n$. This oscillatory behavior was not due to mesoscopic (interference) fluctuations[28]: It remained the same for measurements using different contact configurations and different $D$. The oscillations were even more profound in Hall resistivity $\rho_{xy}(n)$ that reversed its sign many times within the gapped region (Fig. 3b and Supplementary Figure 2b). The observed behavior can be understood as changes in the global interference pattern with varying $n$ and, hence, the Fermi wavelength. This is conceptually similar to Aharonov-Bohm oscillations that are caused by the periodic-in-$B$ phase modulation. The expected main periodicity $\Delta n$ is given[7,8,16] by 4/$A$ which corresponds to one extra electron per AB or BA domain, taking into account the fourfold degeneracy. Figures 3a,b yield characteristic $\Delta n \approx 5\pm0.5\times10^{10}$ cm$^{-2}$ and, hence, $A \approx 0.80\pm0.09\times10^{-10}$ cm$^2$, which agrees well with the area 0.86±0.04 ×10$^{-10}$ cm$^2$ found from the main period of Aharonov-Bohm oscillations. The observed oscillations in $n$ can equally be interpreted as the consecutive filling of electronic minibands formed by the triangular 2D lattice of 1D chiral states as suggested in refs.[14,15] (also, see Supplementary Note 5). The electronic spectrum for each such miniband has both electron and hole like states, which should lead to multiple NPs and sign-changing $\rho_{xy}(n)$. Accompanying oscillations in $\rho_{xx}(n)$ are also expected to appear as the minibands are filled one by one. The number of the observed minibands can be estimated by counting the number of NPs (for example, in Fig. 3b there are about 10 oscillations inside the gapped region ($|n| < 3\times10^{11}$ cm$^{-2}$). As $D$ = 0.5 V nm$^{-1}$ results in $\delta \approx 50$ meV[20], the average width $\varepsilon$ of those minibands is ~ 5 meV.

## Discussion

The reason why the minibands evolve periodically in $B$ (Aharonov-Bohm oscillations) but not fully so in $n$ could be the following. First, the interlayer bias $D$ and finite $n$ remove the degeneracy between AB and BA domains so that 1D states propagating at the opposite sides of AB/BA walls acquire somewhat different Fermi velocities[29]. This should lead to beatings between interference oscillations arising from AB and BA domains. Interference along the longer loops should further contribute to some randomness in the oscillatory behavior. To elucidate the mentioned analogy between oscillations caused by changing the Fermi wavelength and the miniband description, we estimate the energy difference $\Delta E = h v_{DW}/L$ between the states arising from consecutive standing waves in the smallest loop $L = 3\lambda$ where $h$ is the Planck constant and $v_{DW}$ is the drift velocity of the 1D states at AB/BA walls. After taking into account the lifted degeneracy between AB and BA domains, the experimental value for $\Delta E$ is $2\varepsilon \approx 10$ meV, which yields $v_{DW} \approx 10^6$ m s$^{-1}$, comparable to graphene's Dirac velocity. Such a large value of $v_{DW}$ suggests extremely sharp boundaries between AB and BA domains (Supplementary Note 6), which agrees with the strain reconstruction for small-$\theta$ superlattices as found by electron microscopy[12] and infrared nano-imaging[13].

Finally, let us point out some other surprising features of electron transport through the AB/BA domain network. First, the oscillations caused by filling the minibands are rapidly smeared by $T$ and completely disappear above 20 K in both $\rho_{xx}$ and $\rho_{xy}$ (Supplementary Note 4 and Fig. 3c). This is probably expected as the minibands are only 5 meV apart and, therefore, cannot be resolved at such high $T$. In contrast, the Aharonov-Bohm oscillations survive to much higher $T$ (Fig. 2). On one hand, this is perhaps not surprising because of relatively long $L_\phi$ typical for graphene. On the other hand, the robustness seemingly contradicts to the narrow minibands description. To explain this conundrum, we refer to Brown-Zak oscillations[30] that also have their origins in the Aharonov-Bohm effect but appear for superlattices with 2D conductivity rather than in a network of conductive 1D states. Brown-Zak oscillations in graphene superlattices were found to be exceptionally robust and survived above room $T$, despite the thermal smearing covered many minibands[30]. This is because the minibands respond to $B$ in a uniform manner, which was described in terms of changing the average group velocity[30]. A similar description is likely to be applicable to the narrow minibands in MTG and explain the robust Aharonov-Bohm oscillations for the thermally smeared minibands.

Another puzzle is the reversal of the average Hall effect found for our conductive network. This is shown in Fig. 3c where $\rho_{xy}$ is plotted for $D$ = 0 and 0.5 V nm$^{-1}$ at 50 K. Without an interlayer bias (top panel), the Hall response is normal, with positive $\rho_{xy}$ for electrons and negative for holes (for a given $B$ direction). In contrast, as the triangular network was formed by applying the interlayer bias, the Hall effect reversed its sign. The normal behavior recovered only at high $n$, inside the conduction and valence bands. The reversal of the average sign of $\rho_{xy}$ implies that the network's minibands are predominantly hole-like for electron doping and vice versa for hole doping. This observation does not follow from any of the existing models[14,15]. To understand the reversal qualitatively, we evoke an analogy with 1D chiral states in the quantum Hall effect. From a semiclassical perspective, these states can be viewed as skipping orbits and, along the inner boundaries of AB and BA domains, electrons would then circulate in the direction opposite to that for cyclotron orbits, that is, an average effect from such small closed loops would be hole-like. Accordingly, a collection of skipping orbits on a triangular network may result in a hole-like Hall effect. This analogy requires theoretical corroboration in terms of minibands' electronic spectra of MTG.

## Methods

**Device fabrication.** Our MTG samples were encapsulated between two hBN crystals and prepared using the tear-and-stack method (see Supplementary Note 1 for more details). We used few-layer graphite as the back gate to reduce charge inhomogeneity. The final devices were shaped into the Hall bar geometry using electron-beam lithography and reactive-ion etching.

**Electrical measurements.** The measurements were carried out using the standard low-frequency lock-in technique with excitation frequencies of 6 to 30 Hz and currents of typically 10 nA at 0.3 K and 100 nA at higher temperatures.

**Data availability**

The data that support the findings of this study are available from the corresponding author upon reasonable request.

**Acknowledgements:** This work was supported by Graphene Flagship, the Royal Society and Lloyd's Register Foundation. S.V.M. was supported by Russian Science Foundation (grant 17-12-01393), and R.K.K. by an EPSRC fellowship award. S.L. and J.H.E. acknowledge support from the Materials Engineering and Processing program of the National Science Foundation (award CMMI 1538127) and the II–VI Foundation. M. K. was supported by National Research Foundation of Korea (Grant 2018R1A6A3A03010943). A.I.B., W.K., B.T. were supported by Graphene NowNANO Doctoral Training Programme.

**Author Contributions:** S.G.X. and P.K. fabricated the devices. A.I.B. and M.K. carried out electrical measurements with help from R.K.K., S.V.M., D.A.B. and W.K. The results were analyzed by A.I.B and M.K. with help from S.G.X., R.K.K., B.T., V.I.F. and A.K.G. V.I.F. and F.G. provided the presented theory calculations. Hexagonal boron nitride crystals were grown by S.L. and J.H.E. A.K.G. and A.I.B. wrote the manuscript with many contributions from S.G.X., M.K., P.K., I.V.G. and V.I.F. All authors contributed to the discussions.

**Competing Interests:** The authors declare no competing interests.

**SUPPLEMENTARY INFORMATION**

**Supplementary Note 1**
**Device fabrication and electrical measurements.** The studied MTG devices were assembled by the standard dry-transfer[1,2] and tear-and-stack[3,4] techniques as briefly explained below. First, we chose a crystal of hexagonal boron nitride (hBN) that would later serve as the top gate dielectric. The crystal was picked up using a double-layer polymer film that consisted of a thin layer of polypropylene carbonate (PPC) spun onto a polydimethylsiloxane (PDMS) film. Then we used a precision micromanipulator to place this hBN crystal on top of a graphene monolayer (prepared on an oxidized Si substrate) so that hBN covered approximately half of the graphene crystal (Supplementary Figure 1a). Next we slowly peeled the hBN crystal (attached to graphene) off the substrate, which resulted in the graphene crystal teared into two parts. The part remaining on the substrate was rotated by up to 0.1° and then picked up by the graphene-hBN stack. This resulted in MTG on the hBN substrate (right panel of Supplementary Figure 1a). The substrate temperature was kept at ~ 40 °C to reduce thermally-induced strain and, also, to avoid possible spurious rotations induced by annealing. Particular care was taken to avoid any contact between MTG and PPC, which allowed a clean interface between the two graphene layers. Finally, an hBN crystal for the bottom gate dielectric was selected and picked up using the same procedures. The resulting four-layer stack was then released onto a graphite crystal residing on an oxidized Si wafer. This graphite crystal served as a bottom gate electrode. We used regions of the MTG bilayer, which lied outside the graphite gate, to define quasi-1D contacts using the etching recipe reported previously[5]. This was followed by the deposition of Cr (3 nm) and Au (60 nm) to make metallic contacts. Further electron-beam lithography and metal deposition were employed to define the top gate electrode. The latter also served as an etch mask for the final plasma etching and, to this end, had a Hall bar configuration that was accordingly projected onto MTG.

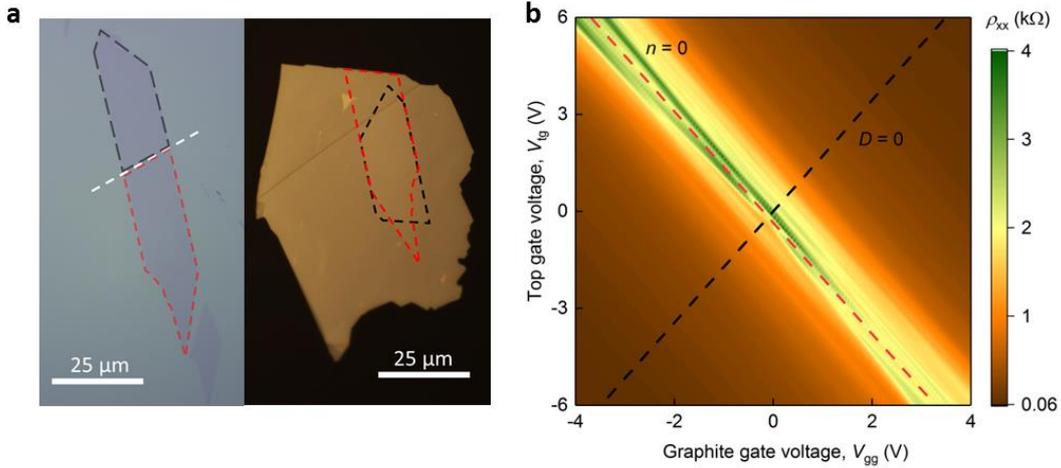

**Supplementary Figure 1 | Tear-and-stack assembly of marginally twisted graphene bilayers and dual-gate characterization. a,** *Optical images illustrating MTG assembly. Left panel: Initial graphene on an oxidized Si wafer. The white line indicates where the tear occurred later. Right panel: Resulting twisted bilayer graphene attached to the top-gate hBN crystal. The black and red dashed curves outline the two teared parts of graphene crystals in both images.* **b,** *Typical map of $\rho_{xx}(V_{tg}, V_{gg})$ for MTG devices at 2 K. The red and black lines show the conditions for zero n and D, respectively. By changing gate voltages to move parallel to these lines allowed measurements under constant n or D.*

The measurements were carried out using the standard low-frequency lock-in techniques with excitation currents below 100 nA, which made heating and nonlinear effects negligible. Most of the data presented in the main text have been taken under the constant displacement field *D*. The dual gated devices allowed us to control its value independently of the total carrier density *n*. The latter is the sum of the densities induced by the top and the bottom gates: $n = \frac{1}{e}(C_{\text{tg}}V_{\text{tg}} + C_{\text{gg}}V_{\text{gg}})$, where *e* is the electron charge, $V_{\text{tg}}$ and $V_{\text{gg}}$ are the top and graphite gate voltages, and $C_{\text{tg}}$ and $C_{\text{gg}}$ are the top and graphite gate capacitances per unit area, respectively. The capacitances were found independently through Hall measurements. The displacement field was calculated as $D = \frac{1}{2\varepsilon_0}(C_{tg}V_{tg} - C_{gg}V_{gg})$, where $\varepsilon_0$ is the vacuum permittivity. To fix $D$, we applied $V_{tg}$ and $V_{gg}$ as illustrated in Supplementary Figure 1b by changing both gate voltages, which allowed us to vary the carrier density $n$ at a constant $D$.

**Supplementary Note 2**
**Determining twist angles.** The actual twist angle $\theta$ between the two graphene layers was determined using two independent methods. First, we used $\rho_{xx}(n)$ and $\rho_{xy}(n)$ measurements to find additional neutrality points (NPs) arising due to the superlattice potential. At zero $D$, they were very close to the main NP, on the steep slopes of the peak in $\rho_{xx}$ (see Supplementary Figure 2a), and we found this procedure unreliable for our devices with extremely small angles. To overcome the difficulty, we applied a finite displacement field, which led to clear NPs in both $\rho_{xx}$ and $\rho_{xy}$ as seen in Fig. 3 and Supplementary Figure 2. This can be attributed to the development of a highly insulating state for the Bernal-stacked AB and BA regions so that they no longer shunted electron transport along AB/BA walls. The higher the displacement field the larger the gap and, hence, more minibands could fit inside.

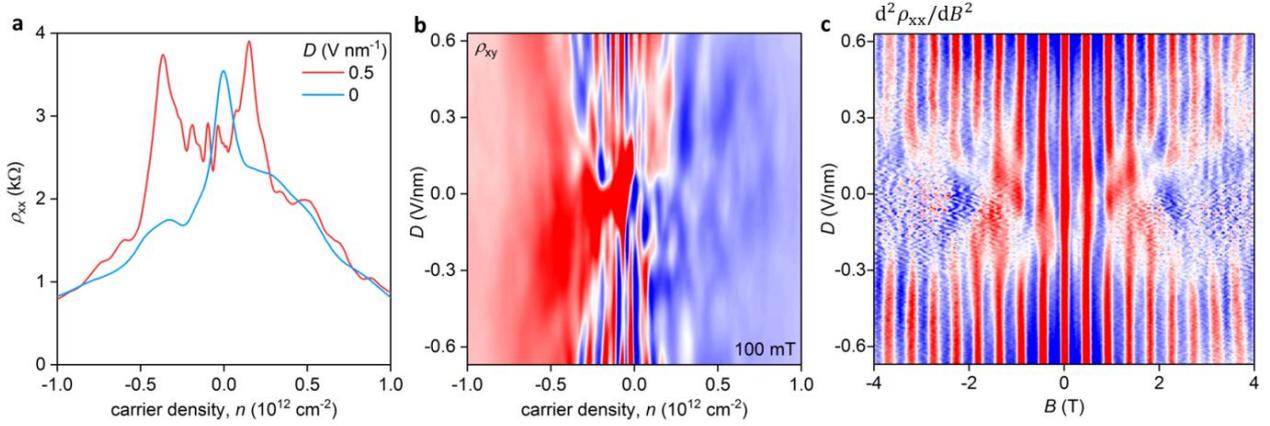

**Supplementary Figure 2 | Multiple neutrality points in the gapped state. a,** $\rho_{xx}$ measured for two displacement fields $D = 0$ and $0.5$ V nm$^{-1}$ at 3 K. **b,** $\rho_{xy}(n, D)$ at $B = 100$ mT and $T = 2$ K. Blue-to-red scale, $\pm 250$ Ohm. **c,** Second derivative of $\rho_{xx}(B)$ as a function of D at 50 K and zero doping. Blue-to-red scale, $\pm 4$ kOhm T$^{-2}$.

As discussed in the main text, the area $A$ of the triangular AB and BA domains is given by $A = 4/\Delta n$ where $\Delta n$ is the distance between the neutrality points. This distance can be found from either $\rho_{xx}$ or $\rho_{xy}$ measurements. For the known area $A$, it is straightforward to find the twist angle

$$\theta = 2\arcsin\left(\sqrt{\frac{\sqrt{3}a^2}{16A}}\right)$$

(see equations for the lattice and superlattice periods in the main text). For the devices discussed in the main text, $\Delta n = 5 \pm 0.5 \times 10^{10}$ cm$^{-2}$ and the above formula yields the twist angle of 0.104° ±0.006°. Another way to determine the twist angle is from the periodicity $\Delta B$ of Aharonov-Bohm oscillations (Fig. 2b and Supplementary Figure 2b). In this approach, the domain area is given by $A = \phi_0/\Delta B$, and from the oscillation period $0.48 \pm 0.02$ T in Fig. 2b of the main text, we find $\theta$ = 0.10 ±0.002°, in agreement with the above estimate using $\Delta n$. Similarly good agreement was found for the other devices. Indeed, for MTG in Supplementary Figure 3a (top panel), $\Delta B = 2.7 \pm 0.2$ T yielded $\theta$ = 0.235±0.01° whereas its $\Delta n = 28 \pm 4 \times 10^{10}$ cm$^{-2}$ yielded $\theta$ = 0.245±0.025°, and the device in Supplementary Figure 3a (bottom) showed $\Delta B = 0.85 \pm 0.05$ T and $\Delta n = 8 \pm 1 \times 10^{10}$ cm$^{-2}$ yielding $\theta$ = 0.133±0.04° and 0.131±0.08°, respectively.

Note that the Aharonov-Bohm oscillations became better developed above a certain displacement field (Supplementary Figure 2c), which is attributed to the fact that 1D electron transport along AB/BA domain walls was no longer electrically shortened by a finite 2D conductivity of the Bernal-stacked regions.

**Supplementary Note 3**
**Further examples of Aharonov-Bohm oscillations.** We fabricated several MTG devices, four of which showing highest homogeneity were studied in detail. All four exhibited pronounced Aharonov-Bohm oscillations under large displacement $D$, and Supplementary Figure 3 shows examples of the magneto-oscillations for two other devices with twist angles of $\sim 0.13°$ and $\sim 0.24°$. At liquid-helium $T$ and in low $B$, both devices showed the first and second harmonics of Aharonov-Bohm oscillations. As the field increased, the Aharonov-Bohm oscillations became overwhelmed by Shubnikov-de Haas oscillations.

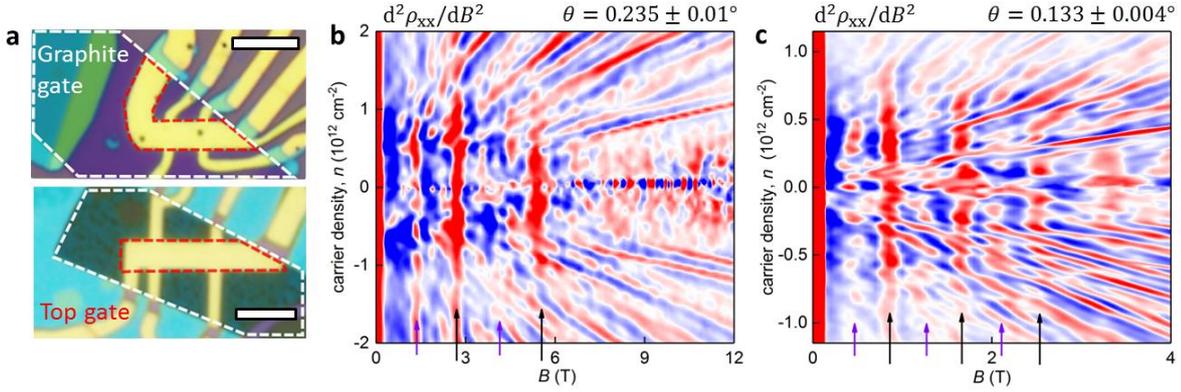

**Supplementary Figure 3 | Aharonov-Bohm oscillations for different twist angles. a**, Optical micrographs of two other devices with $\theta \approx 0.24°$ (top) and $0.13°$ (bottom). Scale bars, 5 μm. **b** and **c**, Second derivative of $\rho_{xx}(B)$ for the MTG in the top and bottom images of (a), respectively. The measurements were carried out at $T$ = 2 K and $D$ = 0.4 and 0.1 V nm$^{-1}$, respectively. Blue-to-red scale: ±6 and ±30 kOhm T$^{-2}$ for (b) and (c), respectively. The black (purple) arrows mark maxima for the first (second) harmonic.

**Supplementary Note 4**
**Temperature dependence in the gapped state.** Despite the large energy gaps $\delta$ were induced by the interlayer bias $D$, all our MTG devices exhibited the metallic behavior at low $n$ inside the expected gapped state (Fig. 1c of the main text). This is further corroborated in Supplementary Figure 4 that shows $\rho_{xx}(n)$ for $D$ = 0.5 V nm$^{-1}$ at different $T$. Although $\delta$ was $\sim$ 50 meV in the Bernal-stacked regions, $\rho_{xx}$ increased with increasing $T$ for $n$ where Aharonov-Bohm oscillations were observed. This density range is indicated by the two dashed lines in Supplementary Figure 4. The figure shows that the additional resistance peaks due to the formation of narrow bands inside the gap became smeared at $T$ of about 20 K. Nonetheless, electron transport due to the in-gap minibands remains seen up to 50 K. This is because charge carriers in the conduction and valence bands led to a lower rate of increase in $\rho_{xx}(T)$, which effectively resulted in the two sharp resistance peaks that can be attributed to the edges of the conduction and valence bands (see the red curve in Supplementary Figure 4). Only at $T$ higher than ~100 K, 2D conductivity inside the Bernal-stacking regions overwhelmed the transport contribution from AB/BA walls, which led to the recovery of the behavior typical for Bernal bilayer graphene[6] with a monotonic decrease in $\rho_{xx}$ with increasing $T$ at low $n$ due to thermally activated carriers. The observed behavior strongly supports the discussed concept of 1D states propagating along AB/BA walls.

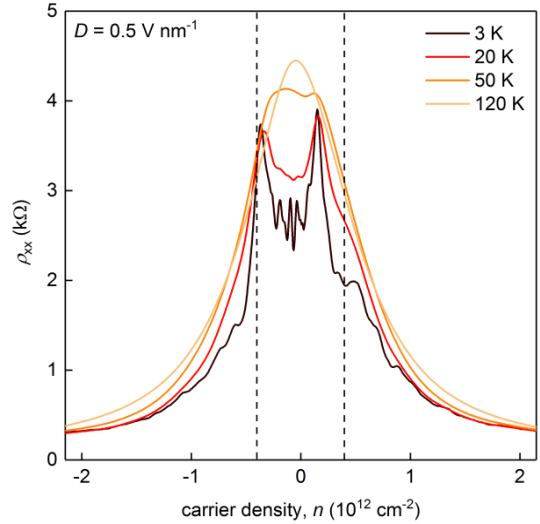

**Supplementary Figure 4 | Resistivity as a function of carrier density at several characteristic temperatures.** $\rho_{xx}(n)$ *was measured at constant D = 0.5 V nm$^{-1}$. The vertical dashed lines indicate the approximate range where Aharonov-Bohm oscillations were observed in this device.*

**Supplementary Note 5**
**Miniband spectrum of the triangular network.** For twisted bilayer graphene, recent theoretical efforts have focused mainly on its properties close to the main magic angle ($\theta \approx 1°$) where the two graphene sheets are assumed to be only rotated with respect to each other and their crystal lattices unperturbed[7–9]. This model is not applicable for marginal angles, $\theta \ll 1°$, where the two graphene layers become considerably strained[10]. The resulting lattice reconstruction was shown to create a rather macroscopic lattice of BLG domains with the conventional staking order and alternating AB and BA regions (Supplementary Figure 5a). The domains are separated by atomically-sharp walls[10].

For such a bilayer system, the interlayer bias opens the standard energy gap $\delta$ within the interior (Bernal-stacked) regions but also gives rise to 1D states propagating along AB/BA domain walls[11] (Supplementary Figure 5). The triangular metallic network can be described by a secondary, miniband spectrum that appears within the energy gap[12]. Supplementary Figure 6a shows this spectrum calculated using the model of ref.[12] where the minibands appear periodically in energy. The model[12] does not take into account changes in the Fermi velocity, which are generally expected to occur near the gap edges (Supplementary Note 6). The latter effect leads to a departure from the periodic behavior, as illustrated in Supplementary Figure 5c.

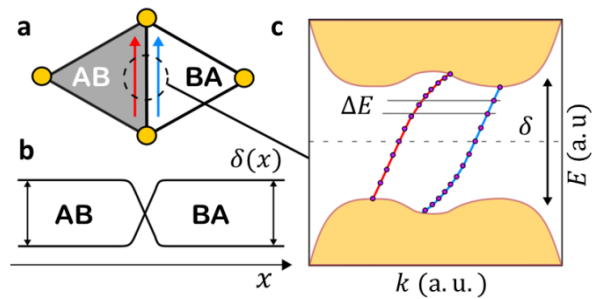

**Supplementary Figure 5 | 1D chiral states at AB/BA domain walls. a**, *Schematic of MTG. White and grey areas: AB and BA domains with the Bernal stacking; yellow circles - regions with AA stacking. The red and blue arrows denote 1D states co-propagating on the opposite sides of the domain wall.* **b,** *Schematic of the gap inversion near the wall between AB and BA domains.* **c,** *Schematic of the domain wall's spectrum*[16]. *The plot shows that as we approach the conduction or valence bands the states on the opposite sides of the*

*domain wall [same color coding as in (a)] acquire different Fermi velocities and, therefore, the corresponding minibands have different spacing (see Supplementary section 6).*

Experimentally, minibands inside the gap should result in multiple peaks in resistivity $\rho_{xx}$ which occur each time the Fermi level moves from one miniband to another crossing the Dirac-like neutrality point where the density of states goes to zero (Supplementary Figure 6b). Also, Hall resistivity $\rho_{xy}$ is expected change its sign. This should happen twice more often than the peaks in $\rho_{xx}$ because the effective mass changes its sign not only at the neutrality points but also at van Hove singularities (Supplementary Figure 6b). The described behaviour is in good agreement with that observed our experiments and shown in Fig. 3 and Supplementary Figure 2.

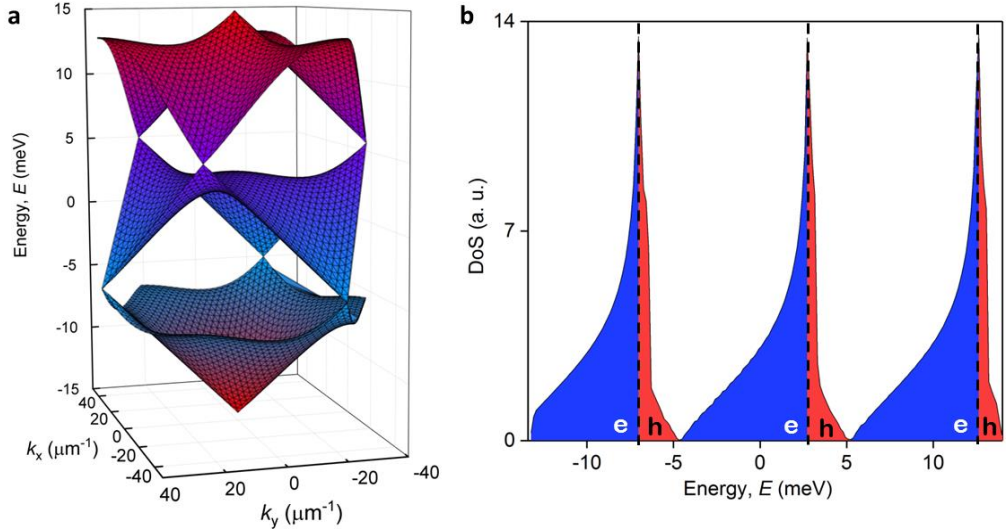

**Supplementary Figure 6 | Miniband spectrum for the triangular network of 1D states. a**, *Spectrum calculated for θ = 0.1° using the model of ref.[12] with the scattering parameter α = 1.1.* **b**, *Density of states for the spectrum in (a). The vertical dashed lines indicate van Hove singularities, whereas the blue and red colored areas mark spectral regions with electron- and hole- like carriers, respectively.*

**Supplementary Note 6**
**Fermi velocity of 1D states at the domain walls.** Recent theoretical studies[11,13–16] confirmed experimentally[17–19] have shown that a domain wall (DW) between oppositely gapped BLG regions (using oppositely oriented *D*), or at the interface between equally gapped AB and BA domains supports two co-propagating chiral 1D states inside the gap for each valley in graphene's band structure (Supplementary Figures 5,6). These states have an almost linear 1D dispersion and propagate in the opposite directions in the two valleys. Electron transport due to these 1D states is protected topologically, unless there are defects that generate intervalley scattering. For a sharp DW having a width $w < v/\sqrt{\gamma_1 \delta}$ (where $\gamma_1$ is the interlayer hopping), the drift velocity $v_{\text{DW}}$ of the 1D states should be of the order of the Dirac velocity $v$ in monolayer graphene, although the exact value is expected to depend on DWs' crystallographic orientation. The co-propagating 1D states are expected[16] to have very close values of $v_{\text{DW}}$, however those diverge for the energies away from the center of the BLG gap and approaching the band edges, as illustrated in Supplementary Figure 5c. Overlaying the two slightly different spectra with a typical energy spacing $\Delta E = v_{\text{DW}} h/3\lambda$ (see the main text) results in a twice denser spectrum with the spacing $\varepsilon \approx v_{\text{DW}} h/6\lambda$. Moreover, differences in the level spacing should lead to beatings in minibands' spectral manifestation (this is noticeable in the data shown in Fig. 3 of the main text). Because each of the standing waves gives rise a different moiré miniband, for a sharp domain wall we expect that the above $\varepsilon$ determines the number of minibands

$$M \approx \delta/(\frac{v_{\text{DW}} h}{6\lambda})$$

which fit inside the gap $\delta$ opened in the Bernal-stacked domains for a given *D*. By counting the number of minibands revealed by the $\rho_{xx}$ and $\rho_{xy}$ oscillations in Fig. 3, we obtained an experimental estimate for the drift velocity as $v_{\text{DW}} \approx v$. Note that, for wider DWs such that $w \gg v/\sqrt{\gamma_1 \delta}$, the 1D states become slower and additional non-topological channels with parabolic dispersions appear. The cumulative effect of the latter two trends should result in a larger number of minibands inside the gap, which can be estimated as $\delta\gamma_1(h\lambda/v)^2$. In the latter case, $v_{DW}$ would be much smaller if estimated from the number of the experimentally observed NPs using the above equation. Therefore, the fact that the estimated $v_{\text{DW}}$ is close to $v$ confirms the general assumption that the lattice reconstruction in marginally twisted BLG transforms it into a network of narrow domain walls between relatively large AB and BA domains with the Bernal stacking.